# The effect of temperature and gas flow on the physical vapour growth of mm-scale rubrene crystals for organic FETs

A.R. Ullah, A.P. Micolich [*], J.W. Cochrane and A.R. Hamilton

School of Physics, University of New South Wales, Sydney NSW 2052, Australia

## ABSTRACT

There has recently been significant interest in rubrene single-crystals grown using physical vapour transport techniques due to their application in high-mobility organic field-effect transistor (OFET) devices. Despite numerous studies of the electrical properties of such crystals, there has only been one study to date focussing on characterising and optimising the crystal growth as a function of the relevant growth parameters. Here we present a study of the dependence of the yield of useful crystals (defined as crystals with at least one dimension of order 1 mm) on the temperature and volume flow of carrier gas used in the physical vapour growth process.

Keywords: rubrene crystal growth, physical vapour transport, organic field-effect transistors, temperature, gas flow.

## 1. INTRODUCTION

Electronic devices based on organic semiconductor materials have been a topic of great interest in recent decades. In part this interest stems from the promise of making cheap, easily produced 'plastic' electronics on flexible plastic substrates,[1,2] but organic semiconductors also promise far superior possibilities for 'chemical tailoring' of the physical properties of a semiconductor (e.g., crystal structure, bandgap, molecular overlap, etc.) than is presently possible with inorganic semiconductors such as GaAs.[3,4] Early studies focussed on conducting polymers such as polyacetylene, polythiophenes, polyphenylene vinylenes and polyanilines.[3,4] However the long chain-length of these conducting polymers severely limits their long-range order, and thereby constrains their electrical mobility and potential for transistor applications.[2,5]

Higher mobilities can be achieved by using shorter chain-length conducting oligomers, such as the acenes (e.g., pentacene, tetracene, etc.), which form highly ordered molecular crystals.[5] Acene-based transistors are commonly produced through the thermal evporation or solution casting of an acene layer on a prefabricated field-effect transistor on a doped Si substrate. However, such acene thin films are typically polycrystalline because crystal growth spreads from various nucleation sites on the substrate surface. Combined with the additional defects caused by the contacts, this presently limits the mobility of acene-on-Si organic field-effect transistor (OFET) devices to ~1 $cm^2$/Vs.[6] An alternative approach leading to higher mobilities involves producing OFETs on high-quality, freestanding organic molecular crystals.[7] Such devices can be made by performing fabrication of the contacts, insulator and gate on the crystal itself,[8,9] or by laminating the crystals onto prefabricated FET structures on polydimethylsiloxane (PDMS)[10-12] or Si substrates.[13,14]

Although tetracene and pentacene have commonly been used for organic crystal transistors for many years, recently the material of choice has become rubrene (5,6,11,12 – tetraphenyltetracene; $C_{42}H_{28}$) for a number of reasons. Firstly, the phenyl side-groups of the rubrene improve the overlap of the electron orbitals of adjacent molecules in the crystal, significantly improving the mobility.[7] Hole mobilities of 40 $cm^2$/Vs have recently been reported for rubrene OFETs,[15] currently the highest value reported for organic transistors. Secondly, high-quality single crystals with areas up to 1 $cm^2$ are readily grown using a relatively simple physical vapour transport technique.[16] Finally, rubrene is significantly cheaper than tetracene and pentacene by a factor of ~5 and ~10, respectively.

Despite there being numerous studies of rubrene OFETs to date,[7-15] as yet there is only one published study exploring the effects of the physical vapour growth conditions on the resulting rubrene crystals.[17] Zeng *et al.* explored the morphology and crystallinity of rubrene crystals using optical microscopy, x-ray diffraction, and scanning electron and

---

atomic force microscopy, as well as the fluorescence spectra of rubrene crystals.[17] However, little data was presented regarding the yield of crystals useful for device applications as a function of the two key growth parameters, temperature and gas flow. In this paper, we present a study of the yield of crystals useful for OFET studies (defined as crystals with at least one dimension > 1mm) as a function of the growth parameters for source temperatures ranging from 280 to 320°C and gas flows ranging from 10 to 60 mL/min.

## 2. EXPERIMENTAL METHODS

The rubrene crystals in this study were grown using the horizontal physical vapour transport growth technique pioneered by Laudise *et al*.[16] A photograph of the furnace we used and its temperature profile are shown in Figs. 1(a) and (b), respectively. The furnace consists of three concentric glass tubes – the inner 19 mm dia. tube is the growth tube, the middle 50 mm dia. tube supports the furnace windings, and the outer 76 mm dia. tube acts as a safety shield to isolate the windings and hot inner parts of the furnace. The inner growth tube is sealed on both ends using ultra-torr fittings, and is connected to a low pressure Ar supply via a controllable flowmeter on the inlet side, and a fume exhaust via a one-way valve and a water bubbler on the outlet side. The furnace windings are made of 0.5 mm dia. nichrome wire, held down with kaolin cement, and spiral wound such that moving along the furnace from the inlet side, the pitch increases rapidly in the first 20% of the length and then decreases slowly over the remaining 80% of the length to produce the temperature gradient shown in Fig. 1(b). The shallow temperature gradient of ~1-2°C/cm is essential to the crystal growth and allows us to separate out impurities with lower sublimation temperatures than rubrene, as shown in the inset to Fig. 1(b). The temperature is controlled using a Jumo iTron PID controller with 0.1°C resolution that is connected to a type K thermocouple placed in between the inner and middle tubes and located at the source zone, and which switches the power supply to the windings via a solid-state relay. Short segments of narrower diameter glass tube are usually placed inside the inner tube (see Fig. 1(b) – inset) to allow the source to be accurately located at the highest temperature point in the furnace and the crystals to be easily recovered and separated from the impurities once growth is complete.

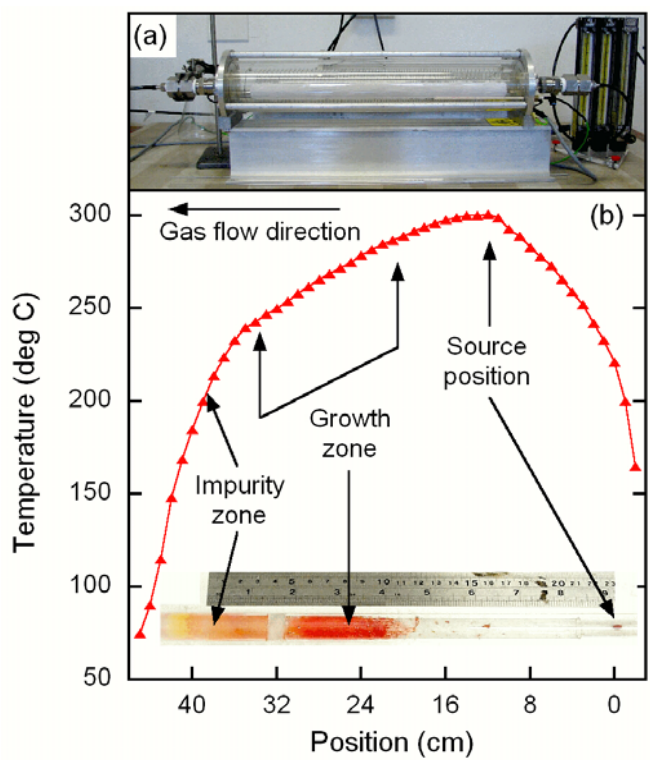

Figure 1: (a) a photograph of the furnace and (b) a graph of temperature as a function of distance from the inlet side of the furnace. Inset to (b) is a photograph of the results of a completed rubrene growth. The source and crystal collection tubes can also be seen.

Crystal growth is performed as follows: 30-50mg of 99% pure rubrene powder (Sigma-Aldrich) is loaded into a 40 mm length of glass tube chosen to be a close fit to the walls of the inner growth tube. The tube is open at both ends and is loaded inside the growth tube such that it sits at the highest temperature point of the furnace windings. Several short lengths (~50-100 mm) of open tube are inserted downstream of the source to facilitate easy removal of the crystals post-growth. The tubes are loaded into the furnace, and the growth tube is purged with a 50mL/min flow of 99.9% pure Ar gas at ~50kPa over-pressure (exhaust is at 1 atm.) for 15-30 minutes to purge the growth tube of air prior to power being supplied to the heating coil. Although other gases such as $N_2$ and $H_2$ can be used,[16] we use Ar because it is inert and readily available. The flow rate for growth is then set and the power is supplied to the windings to set a source temperature of ~ 300°C.

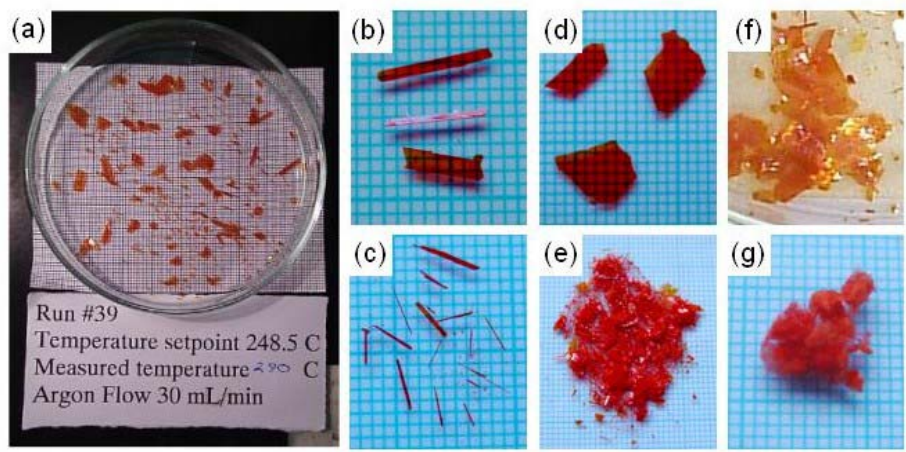

Figure 2: (a) A typical batch of useful crystals collected from a growth run. Crystals are counted into bins based on their area to produce the results in this paper. (b-g) Photographs showing the typical morphologies of crystals obtained by physical vapour growth of rubrene including (b) lath-like crystals, (c) needles, (d) platelets, (e) mix of small needles and powder, (f) thin foil and (g) a cotton-wool-like morphology. All grids shown are 1 mm square.

The growth process is rapid, with small crystallites typically observed downstream within 15-30 minutes of the source temperature reaching ~300°C. The crystals typically deposit where the furnace temperature has fallen to 220-250°C. Growth is usually continued overnight and occasionally for several days (see Sec. 3.5), depending on source temperature, but is always terminated before the source material has been exhausted, as discussed in Sec. 3.1. After the growth is terminated, the growth tube is allowed to cool to room temperature under flowing Ar to avoid oxidation of the crystals, aid the cooling process, and prevent 'suckback' of water from the water-bubbler on the outlet side. The crystals are recovered from the growth tube and the useful crystals, defined as those with at least one crystal dimension greater than 1 mm, are sorted from the large amount of small crystals and crystalline powder that is also obtained (see Fig. 2(e)), and which is recovered for future high-purity growth runs. Care is taken to avoid recovering material from the impurity bands that collect downstream of the rubrene crystals, achieved by using short lengths of collection tubes located inside the inner growth tube. The most common impurity is tetracene, which forms a bright orange impurity band (see Fig. 1(b) – inset). The useful crystals are collected together, as shown in Fig. 2(a), and are counted as a function of area to obtain the results presented in this paper. This process is repeated for the 10 different temperature and gas flow combinations indicated in Fig. 3.

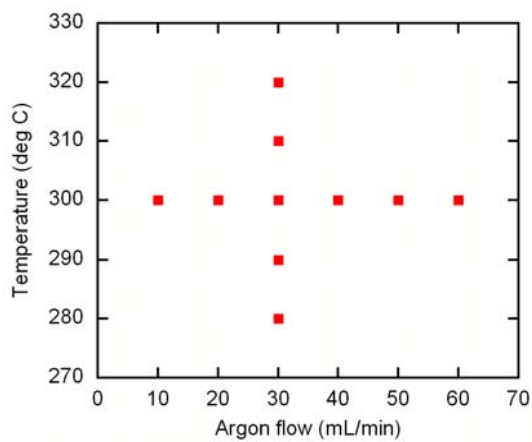

Figure 3: The temperature vs gas flow parameter space explored in this paper. The 10 different temperature and gas flow combinations studied are indicated by the squares.

## 3. RESULTS AND DISCUSSION

### 3.1. Crystal morphologies obtained

Rubrene crystals have an orthorhombic structure[18] with four molecules per unit cell and lattice parameters of $a$ = 14.4 Å, $b$ = 7.2 Å and $c$ = 27.0 Å. A number of different crystal morphologies are obtained by physical vapour growth, the most common ones are shown in Fig. 2(b-g). The lath-like crystals (b) and needles (c) are equally common and are usually longest along the $b$-axis.[7] The platelets (d) are the most useful and in our experience are the most likely to give good electrical properties, but are also the least common morphology. Typically the largest flat facet of the lath-like crystals and platelets is the $ab$-plane.[7] For the results in Sec. 3.3, we count the number of lath-like crystals and platelets that have at least one crystal dimension exceeding 1 mm, after separating them from the other crystal morphologies. We also occasionally find a thin foil morphology, which is typically found at the upstream end of the rubrene growth and for high-growth temperatures and flow rates (see Fig. 2(f)), as well as a cottonwool-like morphology (see Fig. 2(g)) that is typically found downstream of the crystal growth zone and close to the impurity zone. Unlike the platelets, the foil morphology is not flat, it is often curved and is sometimes folded/crumpled, as it appears in Fig. 2(f). Due to its fragility, we will not consider the foil morphology further here and have not yet explored its electrical properties. In addition to the morphologies discussed above and shown in Fig. 2, a typical growth also consists of a large number of small needles and needle-clusters (see Fig. 2(e)), which stick to the glass tube walls, along with a loose crystalline powder, which we suspect are crystallites that form during gas flow and fall once their mass becomes sufficiently large. Together the small needles and powder are often the dominant fraction of the sublimed rubrene by weight, depending on the growth parameters used.

### 3.2. Termination of growth

As mentioned in Sec. 2, the growth is always terminated before the source material is fully sublimed. This is quite important, as the growth of a crystal in this technique is a dynamic process involving the deposition of gaseous rubrene from the source onto the crystal competing against the sublimation of the growing crystal itself due to its elevated temperature. While source material is available, the crystal grows because the deposition rate exceeds the sublimation rate, but once the source material is fully consumed, deposition onto the crystal ceases, and sublimation of the crystal causes it to shrink again. This effect can be clearly observed by taking a time-series of photographs of the growth, where the population center of growing crystals tends to shift slowly downstream with time. If the growth is continued too long after consumption of the source, one ends up with a large number of thick lath-like crystals and needles far downstream and few useful platelet crystals.

### 3.3. The dependence of crystal yield on gas flow

Measurements of the yield of useful crystals were obtained for Ar flows of 10, 20, 30, 40, 50 and 60 mL/min at a constant source temperature of 300°C. Table 1 gives the number of useful crystals, total combined area of the useful crystals, and the area normalised by the sublimed mass of rubrene as a function of flow rate. The sublimed mass of rubrene is obtained by subtracting the mass of rubrene remaining in the source tube after the growth from the mass of rubrene placed in the source tube before growth. Note that this includes the mass of any low sublimation temperature impurities in the rubrene (e.g., tetracene), however we expect these to contribute less than 1% to the sublimed mass. To further characterise the yield of useful crystals, in Fig. 4 (a-f) we plot a histogram of the number of crystals obtained with a given crystal area to the nearest mm$^2$, obtained by counting the crystals over a mm-square grid as shown in Fig. 2(a), for the six different flow rates.

| Flow rate (mL/min) | Number of crystals | Total useful area (mm$^2$) | Normalised yield (mm$^2$/mg) |
|---|---|---|---|
| 10 | 1 | 5 | - |
| 20 | 13 | 63 | 2.0 |
| 30 | 10 | 95 | 1.8 |
| 40 | 27 | 195 | 3.5 |
| 50 | 29 | 365 | 5.8 |
| 60 | 13 | 252 | 4.9 |

Table 1: The number of useful crystals obtained, total area of crystal and total area of crystal normalised by the sublimed mass as a function of flow rate for physical vapour growth of rubrene at a source temperature of 300°C.

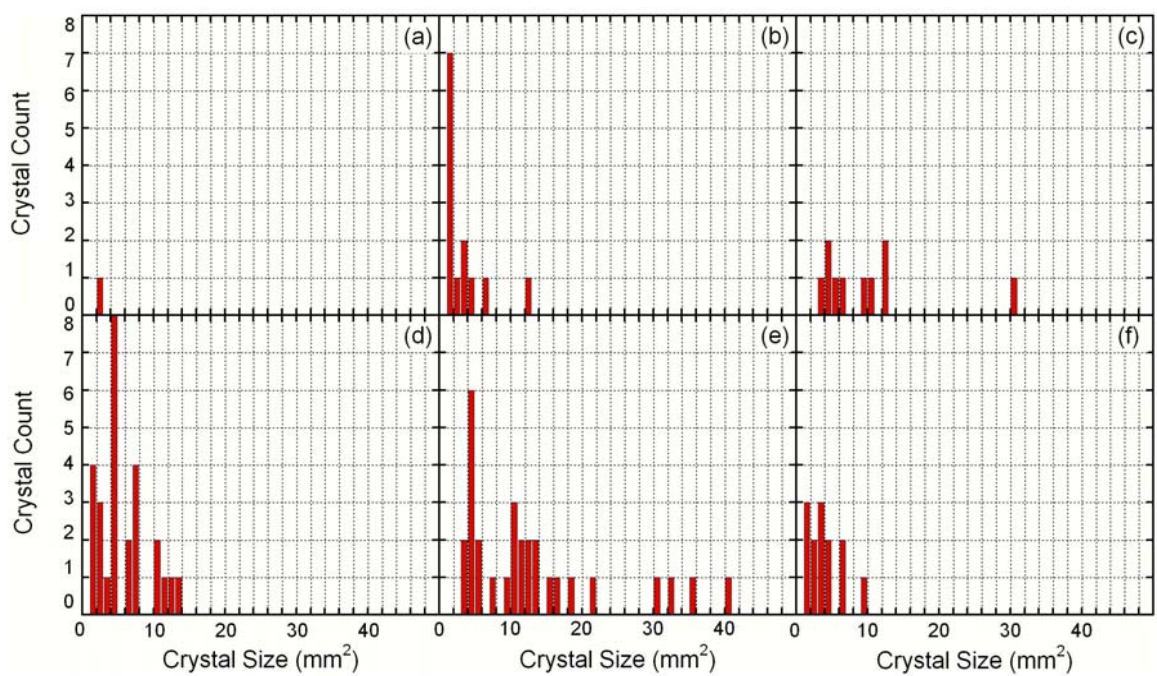

Figure 4: Histograms of the number of useful crystals obtained as a function of crystal area for flow rates of (a) 10, (b) 20, (c) 30, (d) 40, (e) 50 and (f) 60 mL/min for physical vapour growth of rubrene at a source temperature of 300°C.

The yield of useful crystals increases with flow rate up to 50mL/min. This is expected because the gas flow enhances both the sublimation of the source and the transport of any sublimed rubrene down the growth tube. The result is an increased availability of rubrene to deposit on a growing crystal, pushing the balance in favour of growth and leading to favourable conditions for large lath-like crystals and platelets to form (providing the growth is terminated before the source is fully consumed). However, the yield of useful crystals decreases very rapidly for flow rates greater than 50 mL/min. The growth zone is observed to move 47 mm downstream in the growth tube as the flow is increased from 10 to 60 mL/min. Hence the rapid decrease in yield above 50 mL/min is due to the high speed of the flow pushing the growth zone down the furnace to where the temperature falls off rapidly with temperature (see Fig. 1(b)). The rubrene thus obtained is mostly small needle clusters and crystalline powder, and occasionally some rubrene with a cottonwool-like morphology (see Fig. 2(g)), but with very few useful crystals for making OFETs.

### 3.4. The dependence of crystal yield on temperature

Measurements of the yield of useful crystals were obtained for temperatures of 280, 290, 300, 310 and 320 °C at a constant Ar flow rate of 30 mL/min. Table 2 gives the number of useful crystals, total combined area of the useful crystals, and the area normalised by the sublimed mass of rubrene as a function of temperature. To further characterise the yield of useful crystals, in Fig. 5 (a-e) we plot a histogram of the number of crystals obtained with a given crystal area to the nearest mm$^2$ obtained by counting the crystals over a mm-square grid, as shown in Fig. 2(a), for the five different temperatures.

| Source Temperature (°C) | Number of Crystals | Total useful area (mm$^2$) | Normalised yield (mm$^2$/mg) |
|---|---|---|---|
| 280 | 3 | 17 | 0.4 |
| 290 | 5 | 30 | 1.2 |
| 300 | 10 | 95 | 1.8 |
| 310 | 26 | 248 | 2.9 |
| 320 | 34 | 224 | 3.0 |

Table 2: The number of useful crystals obtained, total area of crystal and total area of crystal normalised by the sublimed mass as a function of temperature for physical vapour growth of rubrene with an Ar flow rate of 30 mL/min.

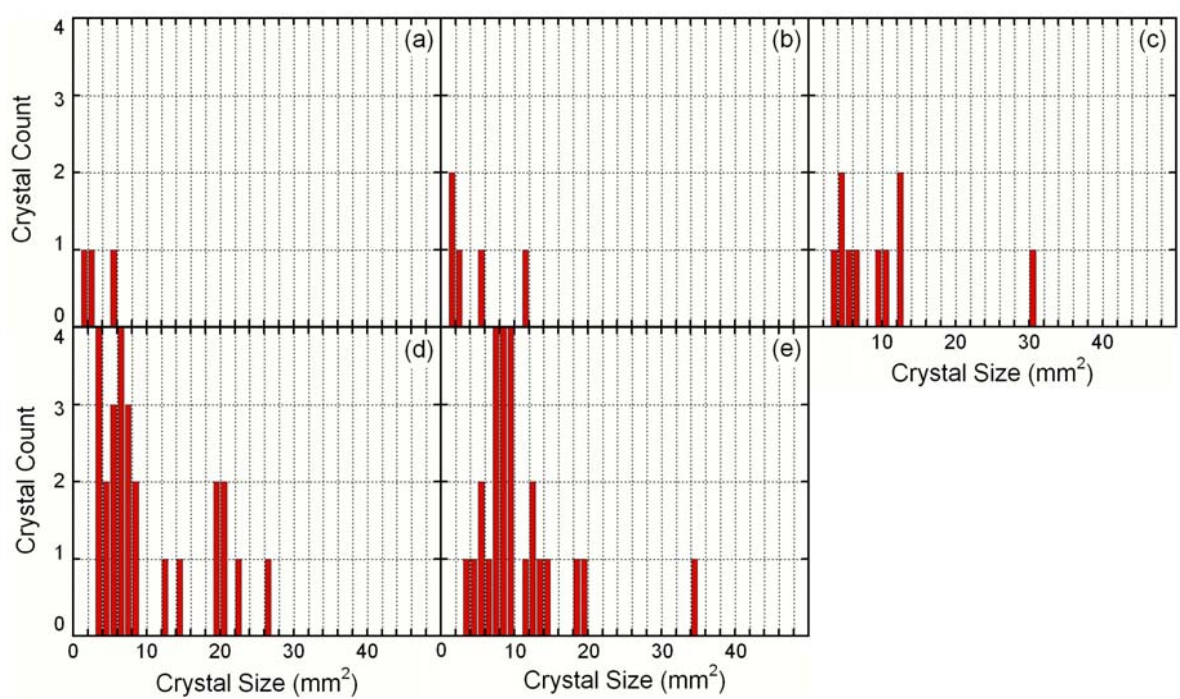

Figure 5: Histograms of the number of useful crystals obtained as a function of crystal area for temperatures of (a) 280, (b) 290, (c) 300, (d) 310 and (e) 320°C for physical vapour growth of rubrene at a flow rate of 30 mL/min.

Generally, as the temperature is increased the number of crystals and the total crystal area increase, as would be expected because the increased temperature accelerates the sublimation process pushing the balance in the favour of the growth of larger crystals. This also has the effect of significantly shortening growth times to a few hours. Increasing the temperature is considerably more effective at reducing growth time than increasing the flow rate at some given temperature. Additionally, increasing the source temperature also pushes the growth zone further down the growth tube, however, the accelerated sublimation discussed above partially compensates for this. Hence a significant yield is still obtained at 320°C, but there is clearly some decrease in the number of very large crystals obtained, as is evident in Fig. 5(e). We expect that further increases in temperature would significantly diminish the yield.

Finally, growth at higher temperatures (e.g. 320°C) results in a lot of foil growth (see Fig. 2(f)) upstream of the normal crystal growth zone. This suggests that growth of the crystals in the *a*-direction (i.e., perpendicular to the fast growth *b*-direction in the needles and lath-like crystals) is driven by temperature, most likely because higher temperatures provide sufficient energy for depositing rubrene molecules to find their way to the edges of a growing platelet. This may also explain why platelets are more commonly found at the higher temperature end of the growth zone, whereas lath-like crystals and needles are more commonly found further downstream. We aim to further explore the link between temperature and the growth of platelets in future studies.

### 3.5. Optimising the yield: temperature vs flow

In Sections 3.3 and 3.4 we showed that both increasing the temperature at constant flow and increasing the flow at constant temperature increase the yield of useful crystals significantly. However, they also shift the growth zone downstream in the furnace. Hence, while at first sight, it may seem logical to increase both temperature and flow (e.g., 310-320°C and 50mL/min), this actually leads to a very low yield because it pushes the growth zone into the steep temperature gradient at the end of the furnace, where growth occurs very quickly giving mostly needles and powder, and little or no impurity separation. So then an interesting question is: Which is better, low temperature and high flow or high temperature and low flow?

We find that low temperature and high flow tends to be better for two reasons. First and foremost, lower temperatures lead to shallower temperature gradients in the furnace (e.g., if the source zone temperature is increased from 300°C to 400°C, the temperature gradient in the growth zone increases by a factor of 2), which leads to better impurity separation. Secondly, for the low temperature/high flow configuration, the growth tends to be slower. This tends to produce better quality crystals, with a larger proportion of useful platelets compared to the high temperature/low flow configuration. In future work, we will aim to make a more quantitative comparison of crystal quality obtained in these two growth configurations, so that more conclusive statements can be made regarding optimisation of the physical vapour growth of rubrene crystals.

## 4. CONCLUSIONS

In conclusion, we have presented a study of the dependence of the yield of useful crystals (defined as crystals with at least one dimension of order 1 mm) on the temperature and volume flow of carrier gas used in the physical vapour growth process. We find that optimum yields occur for lower temperatures and higher flow rates (e.g., 300°C and 50 mL/min) due to the slower growth that occurs compared to using higher temperatures and lower flow rates (e.g, 320°C and 30 mL/min).

## 5. ACKNOWLEDGEMENTS

This work was funded by the Australian Research Council (ARC) under DP0346279, LE0239044, and UNSW internal grants. APM acknowledges the award of an ARC Australian Postdoctoral Fellowship.